\documentclass{aa}
\usepackage{psfig}
\begin{document}

\thesaurus{  02(%
12.03.1;        
02.18.5;        
02.19.2;        
11.09.3;        
11.01.2;        
11.03.4: Coma)  
}

\def\cmb{{\rm cmb}}
\def\a{\alpha}
\def\e{{\rm e}}
\def\o{{0}}
\def\p{{\rm p}}
\def\me{m_{\rm e}}
\def\gh{{\rm gh}}
\def\cl{{\rm cl}}
\def\rg{{\rm rg}}
\def\th{{\rm th}}
\def\rel{{\rm rel}}
\def\jet{{\rm jet}}
\def\power{{\rm power}}
\def\aC{{a_{\rm C}}}
\def\ab{{a_{\rm b}}}
\def\as{{a_{\rm s}}}
\newcommand{\hq}{\hbar}
\newcommand{\epsB}{\varepsilon_{\rm B}}
\newcommand{\epsCMB}{\varepsilon_{\rm cmb}}

\def\eps{\varepsilon}
\def\CR{{\rm cr}}
\def\gal{{\rm gal}}
\def\gas{{\rm gas}}
\def\apj{ApJ}
\def\aap{A\&A}
\def\mnras{MNRAS}

\title{Comptonization of the Cosmic Microwave Background by
Relativistic Plasma} 

\titlerunning{Comptonization of the CMB by Relativistic Plasma}
\authorrunning{T. A.  En{\ss}lin \and C. R. Kaiser} \author{T. A.
En{\ss}lin, C. R. Kaiser} \date{Received 26 January 2000 / Accepted
20 June 2000}  \institute{Max-Planck-Institut f\"{u}r Astrophysik,
Karl-Schwarzschild-Str.1, 85740 Garching, Germany}
\offprints{T.A.E.}  \mail{T.A.E (ensslin@mpa-garching.mpg.de) and
C.R.K. (kaiser@mpa-garching.mpg.de)} \maketitle
\begin{abstract}
We investigate the spectral distortion of the cosmic microwave
background (CMB) caused by relativistic plasma. Within the Thomson
regime, an exact analytic expression for the photon scattering kernel
of a momentum power-law electron distribution is given, which is valid
from the non- to the ultra-relativistic regime. The decrement in the
photon spectrum saturates for electron momenta above $3\,m_\e\,c$ to
that of an optically thick absorber with the optical depth of the
relativistic electrons given by the Thomson limit. Thus, the
ultra-relativistic Sunyaev-Zeldovich (SZ) decrement measures the
electron number and not the energy content. On the other hand, the
relativistic SZ increment at higher frequencies depends strongly on
the spectral shape of the electrons, allowing for investigation of
relativistic electron populations with future instruments.

We calculate the expected Comptonization due to the energy release of
radio galaxies, which we estimate to be $\approx 3\cdot 10^{66} \,{\rm
erg\, Gpc^{-3}}$. We investigate Comptonization from (a) the part of
the released energy which is thermalized and (b) the relativistic,
remnant radio plasma, which may form a second, relativistic phase in
the intergalactic medium, nearly unobservable for present day
instruments (presence of so called `radio ghosts'). We find a thermal
Comptonization parameter due to (a) of $y \approx 10^{-6}$ and (b) an
optical depth of relativistic electrons in old radio plasma of
$\tau_\rel \le 10^{-7}$. If a substantial fraction of the volume of
clusters of galaxies is filled with such old radio plasma the SZ
effect based determination of the Hubble constant is biased to lower
values, if this is not accounted for. Finally, it is shown that a
supra-thermal population of electrons in the Coma cluster would
produce a signature in the Wien-tail of the CMB, which is marginally
detectable with a multifrequency measurement by the Planck satellites.
Such an electron population is expected to exist, since its
bremsstrahlung would explain Coma's recently reported high energy
X-ray excess.

\end{abstract}

\keywords{Cosmology: cosmic microwave background -- Radiation
mechanism: non-thermal -- Scattering -- Galaxies: intergalactic medium
-- Galaxies: active -- Galaxies: clusters: individual: Coma}

\section{Introduction}
\begin{figure*}[tbh]
\begin{center}
\psfig{figure=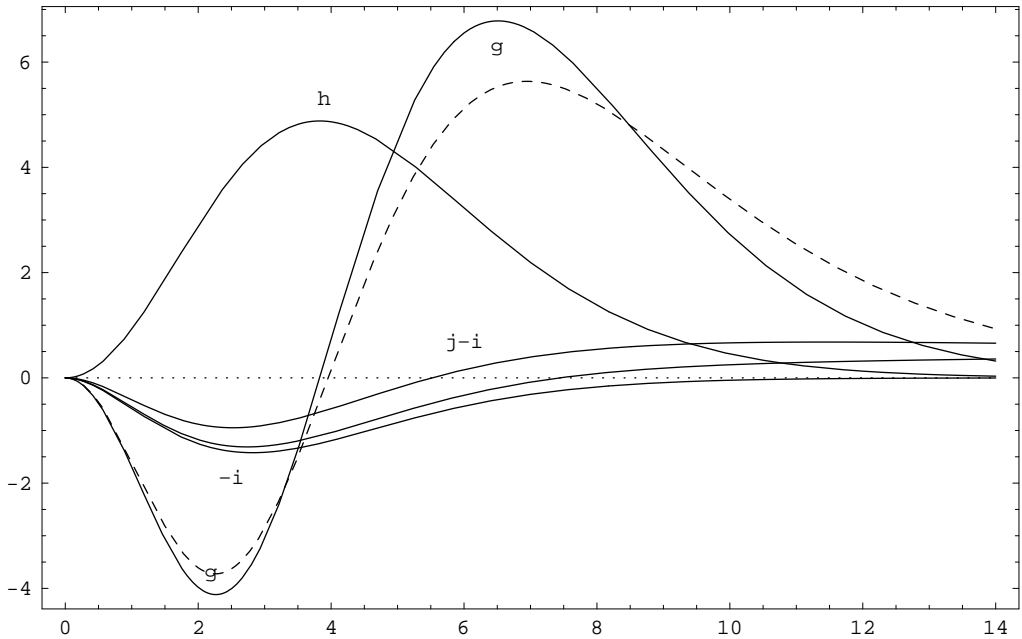,width=0.75 \textwidth,angle=0}
\end{center}
\caption[]{\label{fig:ghij} The functions $g(x)$, $h(x)$, $-i(x)$, and
above the latter $j_\CR(x;\a,p_1,p_2) - i(x)$ for $p_1 =$ 1, 3 (from
top to bottom), $p_2 =1000$, and $\a = 2$. The dashed curve is the
relativistic correct, numerically estimated $\tilde{g}(x)$ for a
thermal plasma with $kT_\e = 15$ keV. }
\end{figure*}

The Sunyaev-Zeldovich (SZ) effect has proven to be an important
tool for cosmology and the study of clusters of galaxies. It measures
the total thermal energy content of the electron population along a
line of sight and does not depend on spectral features of the
underlying electron distributions, as long as they are
non-relativistic. For a review see Sunyaev \& Zeldovich (1980).

Deviations in the spectral shape of the SZ effect occur for higher
electron temperatures or energies, increasing from the mildly, through
the trans- to the ultra-relativistic regime. Approximation for
relativistic corrections to the thermal SZ effect are given in the
literature. For reviews see Rephaeli (1995a), Fargion \& Salis (1998),
Molnar \& Birkinshaw (1999), Birkinshaw (1999), and Sazonov \&
Sunyaev (2000).

The thermal SZ effect (thSZ) is therefore a calorimeter for heat
releasing processes in the universe, such as the gravitational
compression of matter during structure formation, and the energy
output of galaxies. A very violent form of energy release of galaxies
are the outflows of relativistic plasma from active galactic nuclei.
This plasma fills large volumes, the radio lobes, and thereby forms
very low density cavities filled with relativistic particles and
magnetic fields (for evidence for these cavities see: B\"ohringer et
al 1995, Carilli \& Harris 1996, Carilli \& Barthel 1996, Clarke et
al. 1997; McNamara et al. 2000). The replaced inter-galactic medium
(IGM) is heated by this process (e.g. Kaiser \& Alexander
1999\nocite{kaiser99}). It is one goal of this work to estimate the
amount of this energy and its imprint on the CMB (see also En{\ss}lin
et al. 1998b; Yamada et al. 1999).

Also the radio plasma is able to Comptonize the CMB. A search for the
SZ effect in radio lobes was carried out by McKinnon et
al. (1991). After being released from the radio galaxy, the
relativistic electrons with the highest energies cool rapidly due to
synchrotron and inverse Compton (IC) losses, letting the radio lobes
fade away after the AGN stops its activity. Patches of such remnant,
undetectable radio plasma were named `radio ghosts' (En{\ss}lin
1999). However, the less energetic electrons may remain relativistic
in such an environment for cosmological times. Compton scattering off
these electrons removes photons from the CMB Planck spectrum and
scatters them to much higher energies, leading to an SZ effect with a
characteristic spectral signature.

Since radio ghosts are a still speculative ingredient of the IGM, we
discuss different strategies to reveal their presence. We further
discuss their possible influence on the SZ effect in clusters of
galaxies.

The paper is organized as follows. In Sect. \ref{sec:intro} we discuss
the theory of transrelativistic inverse Compton (IC) scattering. In
Sect. \ref{sec:rg} we examine the possible role radio ghosts can have,
and estimate their contribution to the CMB-Comptonization. In
Sect. \ref{sec:cluster} we discuss the possible influence remnant radio
plasma in the ICM can have on the determination of the Hubble
parameter via the SZ effect (Sect. \ref{sec:RPinCL}).  Further SZ
measurements of cluster of galaxies are proposed as a tool to test for
non-thermal electrons in the intra-cluster medium
(Sect. \ref{sec:nthe}).  Conclusions can be found in
Sect. \ref{sec:conclusion}.

We assume an Einstein de Sitter (EdS) cosmology and $H_{\rm o} = 50\,
h_{\rm 50}\,\,{\rm km\,s^{-1}\,Mpc^{-1}}$. All IC calculations are
done in the optically thin limit.

\section{Comptonization of the CMB\label{sec:intro}}
\subsection{Phenomenology of SZ-Effects}

We distinguish here three processes which change the spectral shape of
the CMB by inverse Compton scattering: 
\begin{itemize}
\item the thermal Sunyaev-Zel'dovich effect (thSZ),
\item the kinetic Sunyaev-Zel'dovich effect (kSZ), 
\item up-scattering of photons by relativistic electrons (rSZ).
\end{itemize}
The latter happens whenever a CMB photon gets scattered off a
relativistic electron. An electron with gamma factor $\gamma$
increases the frequency $\nu$ of a scattered photon ($\nu'$) on
average by a factor $\nu'/\nu = \frac{4}{3} \gamma^2 -\frac{1}{3}$. The
photon is therefore scattered to much higher energies and is
effectively removed from the spectral range of the CMB. We consider a
photon to be removed from the CMB, if its energy is increased by about
one order of magnitude or greater, requiring $\gamma >3$. Relativistic
electrons with lower $\gamma$-factors still produce a substantial
energy gain of the photon, which also depopulates the Rayleigh-Jeans
side of the CMB, but these electrons are less effective scatterers
compared to electrons of higher energies. An accurate way to treat
this case is given in Sect. \ref{sec:transrelic}.

Using the usual Kompaneets approximation and neglecting for the moment
those electrons with $\gamma < 3$, the spectral distortion produced by
the three processes under considerations are (see e.g. Rephaeli 1995b)
\begin{equation}
\label{eq:deltai(x)approx}
\delta i(x) = g(x) \,y_\gas - h(x) \, \bar{\beta}_\gas \tau_\gas - i(x) \,
\tau_\rel \,,
\end{equation}
where the CMB-blackbody spectrum is
\begin{equation}
I_\nu = i_0\, i(x) = i_0 \; \frac{x^3}{e^x -1}\, ,
\end{equation}
with $x = h \nu/kT$ and $i_0 = 2\,(kT_\cmb)^3/(hc)^2$. For comparison:
the peak of the spectrum is at $x= 2.82$, and the up-coming Planck
satellite will measure the CMB signal at $x=$ 1.8, 2.5, 3.8, 6.2, 9.6,
and 15 (Bersanelli et al. 1996; Puget 1998; see also
Tab. \ref{tab:planck}).  The thSZ-distortions, which have a spectral
shape given by
\begin{equation}
g(x) = \frac{x^4 \,e^x}{(e^x -1)^2} \left( x\,\frac{e^x + 1}{e^x -1} -
4 \right)\,,
\end{equation}
depend in their strength  on the Comptonization parameter 
\begin{equation}
\label{eq:ygas}
y_\gas = \frac{\sigma_{\rm T}}{\me\, c^2}\, \int\!\! dl\, n_{\e ,\gas}
\,kT_\e \,.
\end{equation}
The line-of-sight integral extends from the last scattering surface of
the cosmic background radiation at $z=1100$ to the observer at $z =0$.
The kSZ-distortions have the spectral shape
\begin{equation}
h(x) = \frac{x^4 \,e^x}{(e^x -1)^2}
\end{equation}
and depend on $\bar{\beta}_\gas$, the average line-of-sight streaming
velocity of the thermal gas ($v_\gas = \beta_\gas \, c $, $\beta_\gas
>0 $ if gas is approaching the observer), and the Thomson optical
depth
\begin{equation}
\label{eq:barbeta}
\bar{\beta}_\gas \, \tau_\gas = \sigma_{\rm T} \int\!\! dl\, n_{\e ,\gas}
\,
\beta_\gas\,.
\end{equation}
Finally, the optical depth of relativistic electrons 
\begin{equation}
\label{eq:yrel}
\tau_\rel = \sigma_{\rm T} \int\!\! dl\, n_{\e,\rel} \,
\end{equation}
determines the fraction of the number of photons, which are
effectively removed from the CMB.

\subsection{Transrelativistic Thomson Scattering\label{sec:transrelic}}

We derive here the exact formulae for the photon redistribution
function of the transrelativistic SZ effect in the Thomson regime,
which can be applied to get the relativistically correct thSZ,
but also the rSZ for any arbitrary isotropic electron
distribution. Such a formula has already been given in the literature
(Rephaeli 1995a), but not in the compact form provided here (see
Eq. (\ref{eq:myP})). We further give the analytically exact photon
redistribution function for a power-law electron distribution.

The change in intensity due to IC-scattering of a blackbody photon
distribution at an isotropic distribution of electrons in the optical
thin limit is given by
\begin{equation}
\label{eq:Deltaiexact}
\delta i(x) = \left(j(x) - i(x)\right) \,\tau \,,
\end{equation}
where $\tau$ is the Thomson optical depth
\begin{equation}
\tau = \sigma_T \,\int \!\! dl \,n_\e\,.
\end{equation}
$i(x) \,\tau$ is the flux scattered to other frequencies, and $j(x)
\,\tau$ is the flux scattered from other frequencies to $x =
h\,\nu/(kT)$.  In Eq. (\ref{eq:deltai(x)approx}) we assumed that
$j(x)\ll i(x)$ in the region of interest ($x<10$) for
ultra-relativistic electrons. In the following we drop this
approximation.

Eq. (\ref{eq:Deltaiexact}) can be re-written to include a $y$-like
parameter
\begin{equation}
\delta i(x) = \tilde{g}(x) \,\tilde{y}\,,
\end{equation}
where
\begin{eqnarray}
\tilde{y} &=& \frac{\sigma_T}{\me\,c^2} \,\int\!\!dl \,n_\e\,k\tilde{T}_\e
\\
k\tilde{T}_\e &=& 
\frac{P_\e}{n_\e} \\ 
\label{eq:tildeg}
\tilde{g}(x) &=& \left(j(x) - i(x)\right)\,\frac{\me \,c^2}{\langle  k
\tilde{T}_\e \rangle }\\
\langle  k\tilde{T}_\e \rangle  &=& \frac{\int\!\!dl \,n_\e \, k
\tilde{T}_\e}{\int\!\!dl \,n_\e}
\end{eqnarray}
with $P_\e$ the electron pressure (see Eq. (\ref{A:eq:Pe1})).  The
definition of the pseudo-temperature $k\tilde{T}_\e$ is equal to the
thermodynamic temperature in the case of a thermal electron
distribution (Eq. (\ref{eq:fth})).

\begin{figure}[t]
\begin{center}
\psfig{figure=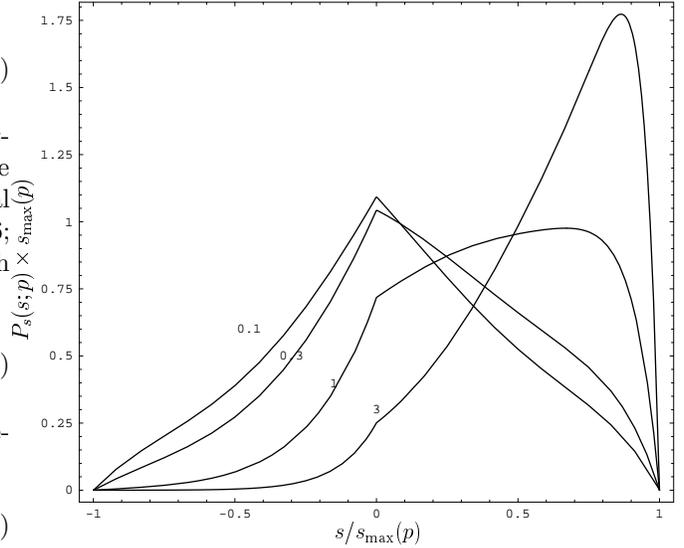,width=0.45 \textwidth,angle=0}
\end{center}
\caption[]{\label{fig:Ps} The frequency redistribution function
$P_s(s)$ plotted in normalized coordinates ($P_s(s;p) \times s_{\rm
max} (p)$ over $s/s_{\rm max}(p)$) for different electron momenta p. From
the left to the right curve is $p= 0.1, 0.3, 1, 3$, yielding $s_{\rm max}
(p) = 0.2, 0.6, 1.8, 3.6$ or $t_{\rm max}(p)= e^{s_{\rm max}(p)} =
1.2, 1.8, 5.8, 38$.}
\end{figure}

\begin{figure}[t]
\begin{center}
\psfig{figure=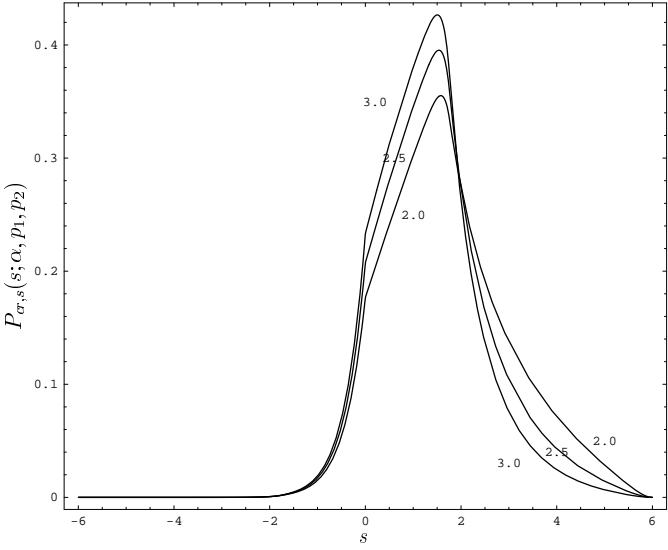,width=0.45 \textwidth,angle=0}
\end{center}
\caption[]{\label{fig:Pscr} The CR-frequency redistribution function
$P_{\CR ,s}(s;\a, p_1,p_2)$ over $s$. $p_1 = 1$, $p_2 = 10$,
and $\a = 3.0, 2.5, 2.0$ from left to right.}
\end{figure}
If one introduces the photon redistribution function $P(t)\,dt$, which
gives the probability that the frequency of a  scattered photon
is shifted by a factor $t$, one can express the scattered spectrum as
\begin{equation}
\label{eq:j} 
j(x) = \int_0^\infty \!\!\!\! dt \, P(t)\, i(x/t)\,,
\end{equation}
where $P(t)$ gives the probability that a photon is scattered to
a frequency $t$ times its original frequency. For a given electron
spectrum $f_\e(p) \,dp$, where $p = \beta_\e\, \gamma_\e$ is the
normalized electron momentum, and where $\int \!\!dp \, f_\e (p) =1$,
the photon redistribution function can be written as
\begin{equation}
\label{eq:P(t)}
P(t) = \int_0^\infty \!\!\!\! dp \, f_\e (p)\, P(t; p)\, .
\end{equation}
$P(t; p)$ is the redistribution function for a momoenergetic electron
distribution.  $P(t; p)$ can be derived following Wright's (1979)
kinematical considerations of the IC scattering in the Thomson regime
($h\,\nu \ll \gamma_{\rm e} \,m_{\rm e}\,c^2$).  Scattering of a
photon from an isotropic photon field has a probability distribution
of the angle $\theta$ between photon and electron direction in the
electron's rest frame given by
\begin{equation}
f(\mu)\,d\mu = [2 \gamma_{\rm e}^4(1-\beta_{\rm e}\mu)^3]^{-1}\,d\mu\,\,,
\end{equation}
where $\mu =\cos\theta$. The frequency shift from $\nu$ to
$\nu '$ of the photon after scattering into an angle $\theta '$  is
\begin{equation}
\label{eq:s}
t = \frac{\nu '}{\nu} = \frac{1+\beta_{\rm e}\mu '}{1-\beta_{\rm
e}\mu} \,.
\end{equation}
The probability distribution of $\mu ' = \cos \theta '$ for a given
$\mu$ is 
\begin{equation}
g(\mu '; \mu)\,d\mu' = \frac{3}{8} [1+\mu^2 \mu'^2 +\frac{1}{2}
(1-\mu^2)(1-\mu'^2)] \,d\mu'
\end{equation}
(Chandrasekhar 1950).
The probability for a shift $t$ given the electron velocity is
\begin{equation}
\label{eq:Pint}
P(t;p) = \int d\mu \,f(\mu)\,g(\mu ';\mu)\,
\left( \frac{\partial \mu '}{\partial t} \right)\,.
\end{equation}
$\mu'(t,\beta_{\rm e},\mu)$ is given by (\ref{eq:s}) and the range of
integration is given by the conditions $-1<\mu<1$ and $-1<\mu'<1$.
The integrand in (\ref{eq:Pint}) is a rational function of $\mu$ and
therefore analytically integrable:
\begin{eqnarray}
\label{eq:myP}
&&{\textstyle P(t;p)} {\textstyle\, =\,} - \frac{3 |1-t| }{32p^6t}
\left[ {\textstyle 1 + (10+8p^2+4p^4)t+t^2} \right] + \\ 
&& \frac{3 (1+t)}{8 p^5 } \left[ \frac{3+3p^2+p^4}{\sqrt{1+p^2}} -
\frac{3+2p^2}{2p} {\textstyle (2\, {\rm arcsinh} (p) - |\ln(t)|)}
\right] \nonumber
\end{eqnarray}
The maximal frequency shift is given by
\begin{equation}
|\ln(t)| \le 2\,{\rm arcsinh}(p)\,,
\end{equation}
and therefore $P(t;p)=0$ for $|\ln(t)| > 2\,{\rm arcsinh}(p)$. Similar
expressions for the photon redistribution function using different
variables can be found in the literature (Rephaeli 1995a; Fargion et
al. 1996; En{\ss}lin \& Biermann 1998; Sazonov \& Sunyaev 2000). We
verified analytically, that $P(t;p)$ has the proper normalization
($\int\!\!dt\, P(t;p) = 1$) and checked numerically that it also
reproduces the known average energy gain of the scattered photons
($\int\!\!dt\, P(t;p)\,(t-1) = \frac{4}{3} \,p^2$).

If one is interested in the large frequency shift ($t \gg 1$) due to
IC-scattering by ultra-relativistic electrons it is convenient to use
the logarithmic frequency shift $s = \ln (t)$.  The photon
redistribution function then reads
\begin{equation}
P_s(s;p) \,ds= P(e^s;p) \, e^s\,ds\,,
\end{equation}
and $P_s(s;p) = 0$ for $|s| > s_{\rm max}(p) = 2\,{\rm
arcsinh}(p)$. $P_s(s;p)$ is plotted in Fig \ref{fig:Ps}.

\begin{figure}[t]
\begin{center}
\psfig{figure=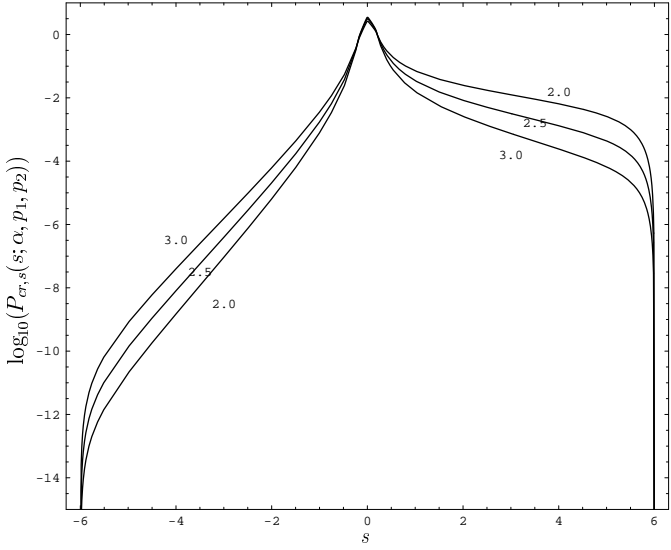,width=0.45 \textwidth,angle=0}
\end{center}
\caption[]{\label{fig:lnPscr} The CR-frequency redistribution function
in logarithmic units $\log_{10} (P_{\CR ,s}(s;\a, p_1,p_2))$ over $s$.
$p_1 = 0.1$, $p_2 = 10$, and $\a = 3.0, 2.5, 2.0$ from left to right.}
\end{figure}

We are interested in three distinct forms of the momentum distribution
of the scattering electrons:
\begin{itemize}
\item 
a thermal electron distribution:
\begin{equation}
\label{eq:fth}
f_{\e,\th}(p;\beta_{\rm th}) = \frac{\beta_{\rm
th}}{K_2(\beta_{\rm th})}\, \, p^2 \,\exp (-\beta_{\rm
th}\,\sqrt{1+p^2})\,,
\end{equation}
with $K_\nu$ denoting the modified Bessel function of the second kind
(Abramowitz \& Stegun 1965\nocite{abramowitz65}), introduced here for
proper normalization, and $\beta_{\rm th} = \me\, c^2/kT_\e$ the
normalized thermal beta-parameter.
\item
a power-law distribution between $p_1$ and $p_2$, as a model for a
cosmic ray (CR) electron population:
\begin{equation}
f_{\e,\CR} (p;\a,p_1,p_2) = \frac{(\a-1)\,  p^{-\a}}{p_1^{1-\a} -
p_2^{1-\a}}
\end{equation}
\item
a thermal distribution with a high-energy power-law tail, as a model
for a thermal population modified by in-situ particle acceleration:
\begin{eqnarray}
\label{eq:fe,pow}
f_{\e,\th\&\CR} (p;\beta_\th,\a,p_1,p_2) =  C_{\rm
e}(\beta_\th,\a,p_1,p_2)\;\times &&\\
\nonumber
\left\{
\begin{array}{ll}
f_{\rm e,th}(p;\beta_\th) &;\;\;\; p\leq p_1\\
f_{\rm e,th}(p_1;\beta_\th) \,(p/p_1)^{-\a} &; \;\;\; p_1<p<p_2\\
0&;\;\;\; p_2<p 
\end{array}
\right\},&& 
\end{eqnarray}
where the normalization parameter $C_{\rm e}(\beta_\th,\a,p_1,p_2)$ is
determined by $\int\!dp\,f_{\e,\th\&\CR} (p;\beta_\th,\a,p_1,p_2) =
1$.
\end{itemize}
The integral in Eq. (\ref{eq:P(t)}) has to be evaluated numerically for
the thermal electron distribution. But the analytical solution for the
power-law case can be expressed as
\begin{small} 
\begin{eqnarray}
&&
P_\CR(t;\a,p_1,p_2) = \frac{\a-1}{p_1^{1-\a} - p_2^{1-\a}} \,\frac{3}{16}
\,( 1+t ) \times \nonumber
\\
&&
\left[
- {\rm B}_{\frac{1}{1+p^2}}\left({\frac{1+\a}{2}},{-\frac{\a}{2}}\right)
\right.  \nonumber\\ && \left. 
- 
{\rm B}_{\frac{1}{1+p^2}}
\left({\frac{3+\a}{2}},-{\frac{2+\a}{2}}\right) 
\,\frac{7+3\,\a}{3+\a} 
\right. \nonumber\\ && \left. 
-
{\rm B}_{\frac{1}{1+p^2}}
\left({\frac{5+\a}{2}},- {\frac{4+\a}{2}}\right) 
\, \frac{12+3\a}{5+\a} 
\right. \\ && \left.
+ p^{-5 - \a}\,
\left\{
\left( \frac{3}{5+\a} + \frac{2\,p^2}{3+\a} \right)
\,\left(2\,{\rm arcsinh}(p) - \left|\ln (t)\right|\right)
\right. \right.  \nonumber\\ && \left. \left.
+
\left| \frac{1-t}{1+t} \right| \,
\left( 
\frac{1+t^2}{2(5+\a) t} + \frac{5}{5+\a} + \frac{4\,p^2}{3+\a}
+\frac{2\,p^4}{1+\a}\right)
\right\}
\right]_{\tilde{p}_1(t)}^{\tilde{p}_2(t)},
\nonumber
\end{eqnarray}
\end{small} 
with ${\tilde{p}_1(t)} = {\rm max}(p_1,\sqrt{t}/2)$ and
${\tilde{p}_2(t)} ={{\rm max}(p_2,\sqrt{t}/2)}$.  Here ${\rm
B}_x(a,b)$ denotes the incomplete beta-function (Abramowitz \& Stegun
1965\nocite{abramowitz65}), and 
\begin{equation}
\label{eq:short}
\left[ f(p)\right]_{p_1}^{p_2} = f(p_2) - f(p_1)\,.
\end{equation}
$P_\CR(t;\a,p_1,p_2) = 0$ if $|\ln(t)| > 2\,{\rm arcsinh}(p_2)$. This
function is plotted in Fig. \ref{fig:Pscr} and \ref{fig:lnPscr}. The
resulting change in a black-body photon spectrum due to Comptonization
by CR-electrons is shown in Fig. \ref{fig:ghij} as the curves labeled
with $j-i$. Note that for a given frequency the influence of the gain
in up-scattered photons of lower initial frequencies, $j_\CR(x)$, on
the spectral distortions are negligible compared to $-i(x)$, which
describes the loss of photons from the CMB of intially this frequency,
as long as $x<10$ and the lower cutoff the electron spectrum $p_1 >
3$. Therefore the rSZ effect can be well approximated as a pure
absorption process of photons in the CMB-spectral range ($x<10$) for
sufficiently relativistic electrons, justifying
Eq. (\ref{eq:deltai(x)approx}).

\begin{figure}[t]
\begin{center}
\psfig{figure=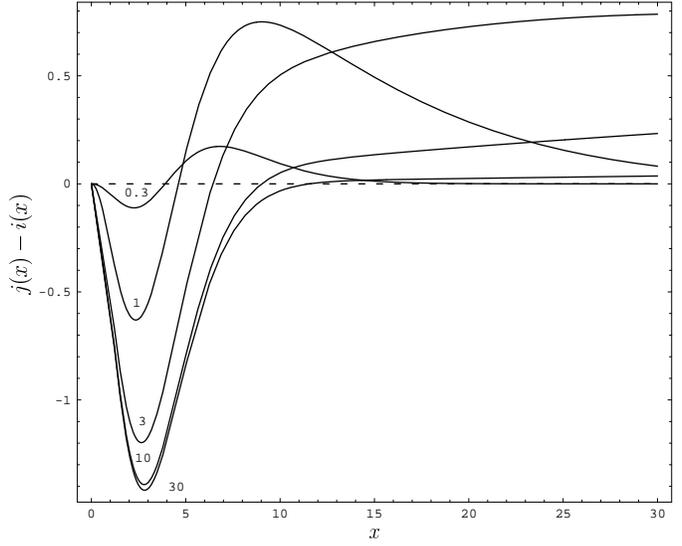,width=0.45 \textwidth,angle=0}
\end{center}
\caption[]{\label{fig:ji} $j(x)-i(x)$ for mono-energetic electron
spectra with $p = $ 0.3, 1, 3, 10, 30. These curves give the spectral
change per electron of this momentum. Note that the absorption feature
saturates to $-i(x)$ for $p \ge 10$.}
\end{figure}

\subsection{\label{sec:emission}Relativistic SZ-Effect in Emission}

\begin{figure}[t]
\begin{center}
\psfig{figure=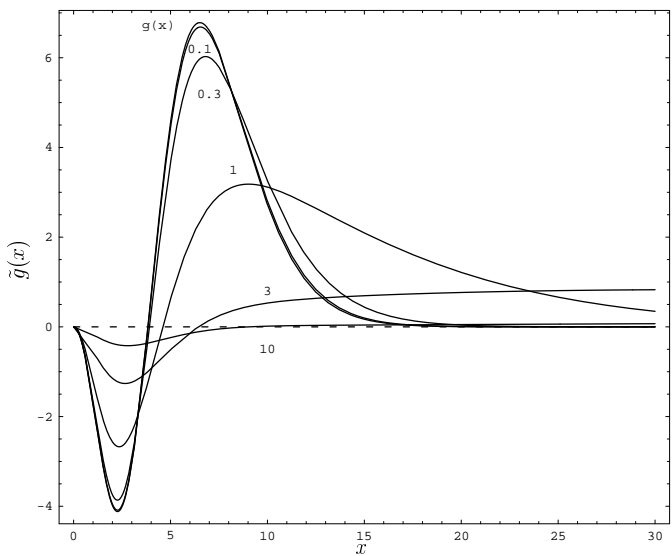,width=0.45 \textwidth,angle=0}
\end{center}
\caption[]{\label{fig:gtild} The relativistically correct
$\tilde{g}(x)$ for mono-energetic electron spectra with $p = $ 0.1,
0.3, 1, 3, 10.  Also the low energy limit $g(x)$ is shown, which is
the curve with the most pronounced extrema. The curves are
proportional to the spectral change per electron pressure.}
\end{figure}

\begin{figure}[t]
\begin{center}
\psfig{figure=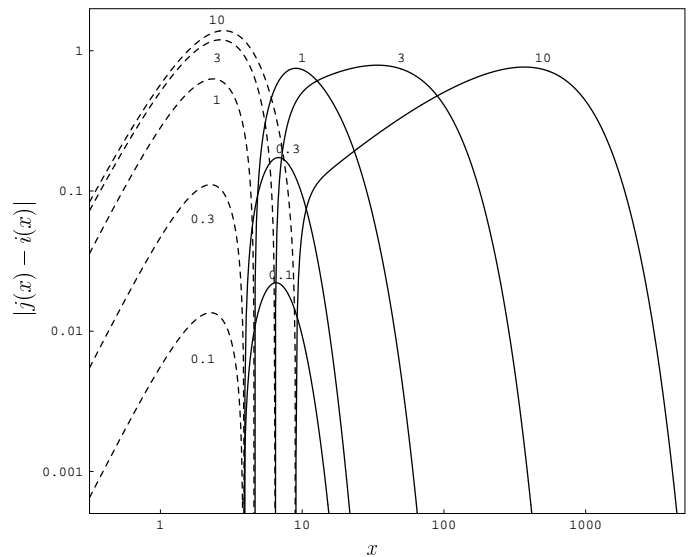,width=0.45 \textwidth,angle=0}
\end{center}
\caption[]{\label{fig:logji} $|j(x) - i(x)|$ for mono-energetic
electron spectra in double logarithmic units. The dashed lines show
absorption, the solid lines emission. The curves are labeled with the
electron momentum.}
\end{figure}

Above the crossover frequency ($x_{\rm cross} \ge 3.83$) the arriving
up-scattered photons outnumber the disappearing ones giving an SZ
increment. This has a different spectral shape in the relativistic
case compared with the classical one. The higher the frequency, the
stronger is the rSZ effect compared to the thSZ effect.  This provides
for a promising opportunity to detect rSZ effect via this
characteristic signature. Fig. \ref{fig:ji} -- \ref{fig:logji}
illustrate this. A direct application of the emission signature of the
rSZ effect can be found in Sect. \ref{sec:nthe}.

\section{\label{sec:rg}Radio Galaxies}
\subsection{Remnant Radio Plasma\label{sec:radiopl}}

The presence of radio plasma produces all three SZ effects discussed
in this paper: The expanding lobes of radio galaxies (RGs) and
radio-loud quasars compress and may shock the ambient IGM, thereby
increasing its thSZ effect. During the inflation phase of the radio
lobes the ambient medium is pushed aside, leading to a possible kSZ
effect. And finally the relativistic electron content of the radio
plasma produces the rSZ effect.

The inflation phase of a RG is short compared to
cosmological times, thus we do not consider the transient kinematic
effect. Furthermore, since gas on both sides of the radio lobes is
accelerated in different directions, giving contributions of opposite
sign but similar strength which should therefore roughly cancel
out. In Sect. \ref{sec:ghostenergy} we argue that the combined kSZ
effect of an isotropic ensemble of flows is practically
indistinguishable from a thSZ effect with the same energy content,
further justifying our simplified treatment.

Radio lobes become unobservable due to adiabatic, IC- and synchrotron
energy losses of the electrons on a time-scale shorter than the
expansion time of the radio lobes which leads to pressure equilibrium
with the surrounding medium (e.g. Kaiser et
al. 2000\nocite{kaiser00}). However, the energy transfered to the IGM
by the expansion of the lobes, and also the relativistic electron
population remain and leave their fingerprints in the CMB spectrum via
Comptonization.

The radio plasma and later radio ghosts will expand or contract until
they reach pressure equilibrium with the surrounding medium. The
pressure of the radio ghost is given by that of the confined
relativistic particles and the magnetic fields, assumed to be in rough
energy equipartition.  Thus the magnetic field energy densities should
be of the order of the thermal energy density of the environment.
Subsonic turbulence in this environment, which has an energy density
below the thermal energy density, is therefore not strong enough to
overcome the magnetic elastic forces of the radio ghost if the
magnetic field in the ghost is contiguous on large scales. In the case
of a magnetic field disjoined on scales smaller than the physical size
of the ghost and/or the turbulence being sonic or super-sonic, which
is e.g. expected in giant merger events of cluster of galaxies, it can
`shred' the ghost into smaller pieces. The size of such pieces will be
comparable to the eddy size of the turbulence. This means, since a
typical turbulent spectrum has less energy density on smaller scales,
that there is a length-scale below which the turbulence is not able to
overcome magnetic forces.  Turbulent erosion of radio ghosts should
stop at this length-scale, leaving small-scale patches of still
unmixed old radio plasma. Similarly, fluid instabilities of
Kelvin-Helmholtz or Rayleigh-Taylor type at the surface of the ghost
will not lead to a complete mixing of the radio plasma with the
surrounding matter.

The relativistic electrons suffer from adiabatic losses during the
inflation phase of the radio plasma, afterwards only from synchrotron
and IC- losses. Coulomb- and bremsstrahlung-losses are negligible in a
tenuous relativistic plasma. The cooling of an ultra-relativistic
electron with momentum $p$ is governed by
\begin{equation} 
\label{eq:cool}
- \frac{dp}{dt} =  \aC + \ab p + \as p^2.
\end{equation}
(Kardashev 1962), where we ignored adiabatic losses or gains due to
volume changes. The coefficient $\aC$, $\ab$, and $\as$ for Coulomb,
bremsstrahlung, and synchrotron/IC losses are (Rephaeli
1979\nocite{rephaeli79}, Blumenthal \& Gould
1970\nocite{blumenthal70})
\begin{eqnarray}
\aC &=& \frac{3}{2} \sigma_{\rm T} \,c\, n_{\rm e} \left( \ln \frac{\me
c^2 p^{1/2}}{ \hq \omega_{\rm p}} +0.22\right),\\
\ab &=& \frac{3\alpha}{\pi} \sigma_{\rm T} \,c\, n_{\rm e} \left(\ln
2p -\frac{1}{3} \right),\\
\label{eq:as}
\as &=& \frac{4}{3} \frac{\sigma_T \, c}{m_\e \,c^2} \, \left( \epsB +
\eps_{\cmb}\right),
\end{eqnarray}
where the electron density of the background gas is $n_{\rm e}$,
$\epsB = B^2/(8\,\pi)$ is
the magnetic field and $\epsCMB$ the CMB photon energy density.  The plasma
frequency is $\omega_{\rm p} = \sqrt{4 \pi e^2 n_{\rm e}/\me}$ and $\alpha$
is the fine-structure constant. In order to judge the importance of
the different terms in Eq. (\ref{eq:cool}) we define the cooling times
\begin{equation}
t_{\rm C} = \frac{p}{a_C} = 22 \, {\rm Gyr}\,\, \left(1 +
0.01\,\ln(p_{1}/n_{\tiny -5})\right)^{-1}\, p_{1}/{n_{-5}},
\end{equation}
\begin{equation}
t_{\rm b} = \frac{1}{a_b} = 8.6\cdot 10^{3} \, {\rm Gyr}\,\, \left(1 +
0.38\,\ln(p_{1})\right)^{-1}\, /{n_{-5}},
\end{equation}
\begin{equation}
t_{\rm s} = \frac{1}{a_s\,p} = 9.8 \, {\rm Gyr}\,\,
 /(\eps_{-11}\,p_{1}),
\end{equation}
where $p_{1} = p/10$, $n_{-5} = n_{\rm e} /({\rm 10^{-5}\, cm^{-3}})$,
and $\eps_{-11} = (\epsB + \eps_{\rm cmb})/({\rm 10^{-11}\,
erg\,cm^{-3}})$. We note, that in the following estimates the radio
plasma is assumed to contain relativistic electrons only above $p =
10$.  From these timescales it becomes obvious that as long as the
electron density within the radio plasma stays below ${\rm 10^{-5}\,
cm^{-3}}$ only synchrotron and IC losses have to be taken into
account. If the radio plasma does not mix on a microscopic scale with
the ambient medium this can be expected. In the following we assume
that this is indeed the case. We note that in the case of a cold
denser gas component in the radio plasma the Coulomb cooling heats the
cold electron and this heat contributes to the thSZ-effect. The
evolution of the relativistic electron distribution function in this
case can be calculated analytically for constant conditions in the
plasma ($B,n_\e,\eps_{\rm cmb} = const$) (En{\ss}lin et
al. 1999)\nocite{ensslin99}.  If Coulomb- and bremsstrahlung-cooling
can be neglected Eq. (\ref{eq:cool}) is solved by
\begin{equation}
p(t) = \left[ \frac{1}{p_0} + \frac{4}{3} \, \frac{\sigma_T \,
c}{m_\e \,c^2} \, \left( \frac{ \langle B^2\rangle }{8\, \pi} + \langle \eps_{\cmb}\rangle \right)
\, t \right]^{-1}\,,
\end{equation}
where the brackets ($\langle ...\rangle $) indicate time-averaged quantities, and
$p_0$ is the momentum at $t=0$. In order to demonstrate that even
under disadvantageous conditions electrons stay relativistic we insert
$p_0 = 10$, $\langle B^2\rangle  = (20 \, \mu{\rm G})^2$, and $\langle \eps_\cmb\rangle  = 11 \,
\eps_{\cmb,{\rm today}}$ corresponding to an time-average over the CMB
energy density between today and $z=2$ in an EdS Universe. The final
electron momentum after $t=10$ Gyr cooling is still $p>3$. Thus,
practically all relativistic electrons in old radio plasma stay
relativistic for cosmological times. Their energies change, but this
is not important for the rSZ effect in the spectral region of the CMB:
the rSZ decrement depends practically only on the number of
relativistic electrons, and not on their energy. This is because all
the up-scattered photons have energies far above the CMB range, so
that their actual energy does not influence measurements within this
range ($x<10$). However, a search for up-scattered photons is a
promising way to detect the rSZ-effect of very low energy relativistic
electrons.

\subsection{Detectability of Radio Ghosts}

Beside the possibility to see a rSZ signature, discussed in this
article, there are a few other possibilities to detect radio ghosts.

Radio ghosts are practically unobservable as long as their electron
population remains at low energies. But if the population is
re-accelerated the ghost becomes radio luminous again. This can happen
when the ghost is dragged into a large-scale shock wave, e.g. in a
merger event of clusters of galaxies or at the accretion shock where
the matter is falling onto a cluster. The emission region is expected
to be irregularly shaped, and should exhibit linear polarization due
to the compression of the magnetic fields in the shock. Such regions
are indeed observed in the periphery of a few clusters of galaxies
and are called `cluster radio relics' (for reviews, see Jaffe
1992\nocite{jaffe92} and Feretti \& Giovannini
1996\nocite{feretti96}). Their properties, such as degree and
direction of polarization, surface luminosity, peripheral position
etc., can be understood within such a scenario (En{\ss}lin et al.
1998a).  The observed rarity of the cluster radio relic phenomena can
be understood in this context: the presence of a shock wave, which
should be quite frequently found in and near clusters of galaxies, is
not sufficient to produce a cluster radio relic, a radio ghost has to
be present at the same location, too.

Another way to detect the presence of radio ghosts is via their
ability to magnetically scatter ultra-high energy (UHE) cosmic rays
(CR) (En{\ss}lin 1999)\nocite{ensslin99b}. Without such scattering the
distribution of sky arrival directions of UHE CR should trace their
source distribution within $\approx 50$ Mpc, due to the limited
distance protons above $3\cdot 10^{19}$ eV can travel without strong
photo-pion energy losses (Greisen 1966, Zatsepin \& Kuzmin 1966). If
the distribution of UHE CR sources follows that of the matter in the
local Universe a non-uniform arrival direction distribution is
expected, contrary to the observations. Medina Tanco \& En{\ss}lin (in
preparation) show that under optimistic assumptions about their
distribution radio ghosts can sufficiently isotropize the UHE CR
arrival directions. Since the UHE CR scattering angle decreases with
particle energy, this scenario can in principle be tested, as soon as
sufficient CR data is available.

\subsection{The Energy Budget of  Radio Ghosts\label{sec:ghostenergy}}

The energy a radio galaxy release into a radio ghost and its
environment during its lifetime is
\begin{equation}
  \label{eq:gh.energy}
  E_\gh = \eps_\gh \, V_\gh + P\, V _\gh
\end{equation}
where $V_\gh$ is the volume occupied by the ghost, $P$ the pressure
inside and outside the ghost and
\begin{equation}
\eps_\gh =  \eps_\e + \eps_\p + \eps_B  = (1+k_p+k_B) \eps_\e
\end{equation}
is the energy density in the remnant radio plasma, split into its
leptonic ($\eps_\e$), hadronic ($\eps_\p$) and magnetic ($\eps_B$)
part. Unfortunately, the ratios between these energy densities are
still undetermined. The parameters $k_p = \eps_p/\eps_e$ and
$k_B=\eps_B/\eps_e$ take account of this.

The radio plasma is expected to follow a relativistic equation of
state ($P = \frac{1}{3} \eps_\gh$; note that isotropically oriented
magnetic fields also follow a relativistic equation of state). The
fraction of $E_\gh$ stored as volume work in the ambient medium is
therefore
\begin{equation}
\label{eq:DeltaEth}
\Delta E_\th = P\, V_\gh = \frac{1}{4} \,E_\gh\,.
\end{equation}
This thermal energy produces a thSZ effect in addition to that
expected from the heat in the IGM due to structure formation. It seems
to be difficult to follow this energy in an expanding Universe with
ongoing structure formation. But it can be argued that this energy is
released into gravitationally bound structures (clusters and filaments
of galaxies). Filaments expand due to the Hubble flow, but the matter
within a given filament flows into the next galaxy cluster and gets
compressed thereby. This means, that on a longer time-scale the extra
heat from RG is expected to be confined in clusters of galaxies and
does not suffer from adiabatic losses.

In fact, part of the energy injected into the environment might also
be stored gravitationally whenever gas is pushed to a higher
gravitational potential or transformed to kinetic energy of gas
flows. Gravitational energy is converted to kinetic energy, whenever
the accelerated gas flows into the next potential well. From
Eqs. (\ref{eq:deltai(x)approx}) and (\ref{eq:barbeta}) one would
conclude that the kSZ effect is zero for an average of isotropically
oriented flows. But Eq. (\ref{eq:deltai(x)approx}) is only an
approximation. A better description of the average kSZ effect for an
isotropic ensemble of flows can be gained by the following argument:
the individual electron velocities of all flows measured in the
CMB-rest frame can be combined into a single momentum space
distribution function, which is of course isotropic. This function can
be used in order to calculate the Comptonization with the formalism
described in Sect. \ref{sec:transrelic}, since the latter does not
depend on the position of an electron along the line of sight in the
optically thin limit. The kinetic energies are non-relativistic,
so that the spectral changes are well described by the function $g(x)$
times the total kinetic energy of the electrons (see
Fig. \ref{fig:gtild}). Thus the Comptonization is independent of the
exact spectral shape of the electron spectrum or the distribution of
flow velocities.  Therefore we assume that all of the energy, which
the inflating radio lobes give to their environment, contributes to
the thSZ effect. A similar argumentation for gaussian, isotropic,
random velocity fields was given in Zeldovich et al. (1972).

We approximate the relativistic electron spectrum by a single
power-law (Eq. (\ref{eq:fe,pow})), so that the energy density $\eps_\e$
and the number density $n_{\e,\CR}$ of the relativistic electron of a
ghost are related via the average kinetic energy of the electrons
($\langle \gamma_\e -1\rangle \,m_\e\,c^2$):
\begin{equation}
\label{eq:epse1}
\eps_\e =  n_{\e,\CR}\,\langle \gamma_\e -1\rangle  \, m_\e\,c^2 
\approx \frac{\alpha -1}{\alpha -2} \,{n_{\e,\CR} \,m_\e\, c^2\, p_1}\,.
\end{equation}
The approximation assumes the particles to be
ultra-relativistic ($1\ll p_1 < p_2$) and  the distribution to be
dominated by the lower end ($p_1 \ll p_2$ and $\alpha >2$). In the
case that the distribution has a trans- or non-relativistic part the
exact formulae for energy density, pressure and adiabatic index can be
found in Appendix A.

If the ultra-relativistic approximation is valid and the low energy
electrons dominate the spectrum, we can write:
\begin{equation}\label{eq:epsghtone}
n_{\e,\CR} = 
 \frac{\alpha -2}{\alpha -1} \,\frac{\eps_\gh}{(1+ k_\p +
 k_B)\, m_\e\, c^2\, p_1}\,.
\end{equation}
All the quantities on the rhs have to be inserted for the same stage
of the radio plasma evolution in order to give a consistent
result. The best observational constraints for these quantities are of
course given for the moment of injection, when the radio plasma is
visible.

\subsection{CMB-Distortions\label{sec:cmbdist}}
We estimate now the Comptonization due to energy from radio galaxies
stored in different forms into the IGM. We have to calculate
the integrals in Eq. (\ref{eq:ygas}) and (\ref{eq:yrel}) in a cosmological
context. We need to know the distribution of injected heat and
relativistic electrons. Since both quantities (injected thermal heat
in the IGM and number of relativistic electrons in the radio plasma)
are accumulative in the sense, that radio galaxies produce them, but
cooling mechanisms are insufficient to remove them. For the thermal
energy this is clear, since except for cooling flows in the very center
of some cluster of galaxies the cooling time exceeds the Hubble
time. For the number density of relativistic electrons this follows
from the fact that the dominant cooling mechanism in a tenuous
relativistic plasma are synchrotron- and IC-cooling, which become
inefficient for low energy electrons (see Sec \ref{sec:radiopl}).

The integrands in Eq. (\ref{eq:ygas}) and (\ref{eq:yrel}) are
proportional to the amount of remnant radio plasma at a given epoch
(e.g. measured by the number density of relativistic electrons
$n_{\e,\rel}$). Since, as we argued, radio plasma is conserved, its
electron density is given by
\begin{equation}
n_{\e,\rel}(t) = \int_{0}^t \!\!\! dt' \,\dot{n}_{\e,\rel}(t')\,,
\end{equation}
where $\dot{n}_{\e,\rel}(t)$ is the source electron density of radio
plasma at time $t$ in comoving coordinates.  The optical depth of
relativistic plasma can then be written as
\begin{equation}
\tau_\rel = \sigma_{\rm T} \, c\, \int_0^{z_{\rm max}} \!\!\!\!
dz\,\frac{dt}{dz} \dot{n}_{\e,\rel}(z)\, \int_0^z \!\!\!\!
dz'\,\frac{dt}{dz'} \,(1+z')^3\,,
\end{equation}
where $t(z)$ is the time between today and redshift $z$, and $z_{\rm
max}$ is the maximal redshift of injection. We use $z_{\rm max} = 4$,
which is sufficiently low so that cooling effects can be neglected.
Since $\dot{n}_{\e,\rel}(z)$, the source density of relativistic
electrons, is only poorly constrained observationally, we use
$Q_\jet(t)$, the source density of energy from RG instead. For
convenience, we define
\begin{equation}
\omega_\gh = \frac{\sigma_{\rm T} \, c}{m_\e\, c^2} \, \int_0^{z_{\rm
max}}
\!\!\!\!  dz\,\frac{dt}{dz} \dot{Q}_{\jet}(z)\, \int_0^z \!\!\!\!
dz'\,\frac{dt}{dz'} \,(1+z')^3\,,
\end{equation}
and the total amount of energy release through radio plasma:
\begin{equation}
\bar{\epsilon}_\gh = \int_0^{z_{\rm max}} 
\!\!\!\!  dz\,\frac{dt}{dz} \dot{Q}_{\jet}(z)\,.
\end{equation}
For an ultra-relativistic and steep electron population inside ghosts
the approximation Eq. (\ref{eq:epsghtone}) gives
\begin{equation}
\tau_\gh = \frac{3}{4}\,\frac{\alpha -2}{\alpha -1}
\,\frac{\omega_\gh}{(1+k_\p + k_B)\, p_1}\,.
\end{equation}
A similar expression gives the additional thSZ effect due to IGM
heating by expanding radio lobes:
\begin{equation}
\label{eq:ygh}
y_\gh = \frac{\omega_\gh}{12}\,.
\end{equation}
This follows from Eq. (\ref{eq:DeltaEth}) if we assume that the IGM
protons get half of the heat energy.

The optical depth of radio ghosts can be written as
\begin{equation}
\label{eq:ytotau}
\tau_\gh = \frac{\alpha -2}{\alpha -1} \,\frac{9\,y_\gh/p_1}{(1+ k_\p +
k_B)} = 
\frac{y_\gh}{10}\, \frac{3}{1+ k_\p + k_B}\, \frac{10}{p_1}\,.
\end{equation}
Here, we inserted $\alpha = 2.5$. If we assume protons, electrons and
magnetic fields to have all the same energy density, and insert a
speculative lower electron cutoff of $p_1 =10$, this equation suggests
that $y_\gh \gg  \tau_\gh$. This means that the thSZ
decrement 
due to the thermal IGM heating by inflating radio lobes 
always dominates over the rSZ decrement 
of the relativistic electron population in these radio lobes
since $|g(x)| \gg  |i(x)|$ in the Rayleight-Jeans regime (see
Fig. \ref{fig:ghij} and Eq. (\ref{eq:deltai(x)approx}) for details).
The thSZ increment in the Wien-regime decreases exponential and
therefore much faster than the rSZ increment of a power-law electron
population. Therefore the rSZ effect of radio plasma dominates over
its induced thSZ effect at sufficiently high frequencies as long as
the electron population stays relativistic.

The total jet power of radio galaxies per comoving volume can be
derived from the radio luminosity function of RG with the additional
assumption that there is a unique relation between radio luminosity
and jet power. This is not strictly correct, since it is known that
for the largest part of the observed lifetime of RGs they should
exhibit some luminosity evolution, even in models with constant jet
power (for the case of powerful RG of type FRII see e.g. Begelman \&
Cioffi (1989), Falle (1991), Kaiser \& Alexander (1997), Kaiser et
al. (1997), Daly (1999)). But this evolution is `modest' (roughly an
order of magnitude) and its influence should produce some scatter in
an empirical derived relation, but not a large systematic
effect. Similarly we neglect the influence of the density of the gas
in the radio source environments on its radio luminosity. Again we
expect this only to increase the scatter about the empirical function
we will use in the following to relate radio luminosity and jet power.

En{\ss}lin et al. (1997) have fit a power-law to the jet power --
radio luminosity relation derived by Rawlings \& Saunders (1991) for
a sample of radio galaxies, which includes both, the most powerful
radio galaxies of type FRII and also the somewhat less luminous FRI
objects. Their results are based on minimum energy arguments and age
estimates. The real energy of radio plasma can easily be much higher
than the minimal energy estimate by some factor $f_\power >1 $ due to
the presence of relativistic protons, low energy electrons or
deviations from equipartition between particle and field energy
densities. A rough estimate of $f_\power$ can be derived from
observations of radio lobes embedded in the intra-cluster medium (ICM)
of clusters of galaxies. These observations show a discrepancy of the
thermal ICM-pressure to the pressure in the radio plasma following
minimal energy arguments of a factor of $5 - 10$, even if projection
effects are taken into account (e.g. Feretti et al. 1992). Since also
a filling factor smaller than unity of the radio plasma in the radio
lobes can mimic a higher energy density we chose $f_\power = 3$ in
order to be conservative.

The jet power -- radio luminosity correlation at $\nu = 2.7 $ GHz is
\begin{equation}
q_{\rm jet}(L_\nu) = a_\nu  \,(L_{\nu}/({\rm Watt\,
Hz^{-1}}\,h_{50}^{-2}))^{b_\nu}\, f_\power
\end{equation}
for which $b_\nu =0.82 \pm 0.07$, $\log_{10} (a_\nu / {\rm erg\,
s^{-1}}\, h_{50}^{-2}) = {45.28 - 26.22 \cdot b_\nu \pm 0.18}$
(En{\ss}lin et al. 1997). This relation allows to integrate the
observed radio luminosity function $n_\nu (L_\nu,t)$ in order to get
the total jet power
\begin{equation}
\label{eq:Qq}
Q_\jet(z) = \int_{L_{\rm min}}^{L_{\rm max}} \!\!\!\! \!\!\!\!
\!\!\!\! dL_\nu \, n_\nu (L_\nu,z)\, q_\jet(L_\nu)\,.
\end{equation}
We use $L_{\rm min} = 10^{23}\,{\rm Watt\, Hz^{-1}}$, which is just
above the region of the RLF which is dominated by starburst galaxies,
and $L_{\rm max} = 10^{29}\,{\rm Watt\, Hz^{-1}}$.  We adopt different
radio-luminosity functions parameterized by Dunlop \& Peacock (1990)
and integrate Eq. (\ref{eq:Qq}) in an EdS-cosmology up to redshift
$z_{\rm max} = 4$.  Strictly, the expressions for $n_\nu (L_\nu,t)$
used here only apply to those members of the radio source population
with steep radio spectra (spectral index $-0.7$ or less). However, the
space density of the flat spectrum population, which is thought to
consist mainly of FRI-type objects, is at least a factor 10 smaller at
any given redshift (Dunlop \& Peacock 1990). Therefore we can safely
neglect their contribution to the overall radio luminosity function.
We further calculate the ghost distribution function for $b_\nu = 0.7,
0.82, 1$ in order to show the dependence on the uncertainties.

\begin{table}
\begin{tabular}{|cl|cc|}
\hline
RLF & $b_\nu$ & $\bar{\epsilon}_\gh$ & $y_\gh$ \\
\hline
& & $10^{66}\,{\rm erg\, Gpc^{-3}}$ &  $10^{-6}$ \\
\hline
PLE  & 0.7  & 2.80 &  1.51 \\
PLE  & 0.82 & 2.19 &  1.23 \\
PLE  & 1.0  & 1.91 &  1.15 \\
RLF2 & 0.7  & 4.86 &  1.94 \\
RLF2 & 0.82 & 3.04 &  1.40 \\
RLF2 & 1.0  & 2.04 &  1.18 \\
\hline
\end{tabular}
\caption[]{\label{tab:totaljetpower} Cosmological energy output of
radio galaxies for two radio luminosity functions of Dunlop \& Peacock
(1990) and expected Comptonization parameter (Eq. (\ref{eq:ygh})).
PLE denotes the pure luminosity evolution and RLF2 is the second
free-form model of Dunlop \& Peacock. The optical depth $\tau_\gh$ of
the relativistic electrons in ghosts is not well determined since it
depends on several unknown parameters.  Eq. (\ref{eq:ytotau}) should be
used in order to translate $y_\gh$ to $\tau_\gh$.  If the internal
energy of the radio plasma is fully thermalized, then $y_\gh$
increases by a factor of up to 4.}
\end{table}
Results of the integration of the different RLFs and jet power-radio
luminosity correlations are given in Tab. \ref{tab:totaljetpower}. The
energy input in form of radio plasma is roughly $\bar{\epsilon}_\gh =
3\cdot 10^{66} \,{\rm erg\, Gpc^{-3}}\,(f_{\rm power}/3)$. 

Note that the X-ray background, which is believed to be dominated by
AGN emission, corresponds to an injection energy of $10^{67-68} \,{\rm
erg\, Gpc^{-3}}$ in comoving coordinates (Soltan 1982; Chokshi \&
Turner 1992; Fabian \& Iwasawa 1999; Fabian 1999). Either the X-ray
energy losses of AGNs exceed the radio plasma release, or if X-ray
power is comparable to the jetpower then the latter is strongly
underestimated here and $f_\power \approx 30$ would be more
realistic. The resulting Comptonization would be comparable to the
present-day upper limit of $y<1.5 \cdot 10^{-5}$ (Fixsen et al. 1996).
The extra-galactic radio background has an energy density of $\approx
5\cdot 10^{63}\,{\rm erg\,Gpc^{-3}}$ (Longair \& Sunyaev 1971), which
is several orders of magnitude below the energy density in the X-ray
background and the expected ghost energy density. This indicates that
radio emission is a very inefficient mechanism to extract energy from
radio plasma.

It is interesting to note that the X-ray background predicts a mass
density in AGN black holes of $(1.4 ... 2.2) \cdot
10^{14}\,M_\odot\,{\rm Gpc^{-3}}$ for a
mass-to-light-conversion-efficiency of AGNs of 0.1. This is consistent
with the central black hole mass density derived from the observed
black-hole-to-galactic-bulge-mass ratio and the observed masses of
galaxies (Faber et al. 1997).

If the total energy of the radio lobes would be thermalized, then no
rSZ-effect would result, but the contribution to the thSZ effect would
increase by a factor of 4, shifting it closer to the present upper
limit.

\subsection{Discussion}

The expected thermal Comptonization due to heating of the environment
of RG is of the order of $y = 1.4\cdot 10^{-6}\, (f_{\rm power}/3)$,
if it is assumed that radio ghosts do not thermalize their internal
energy, otherwise up to a factor of 4 higher. $f_{\rm power}$ is the
ratio of true energy over equipartition energy in radio lobes.

There are two attempts in the literature to estimate the thSZ effect
due to energy release from RGs. 

En{\ss}lin et al. (1998b) give $y \le 1.0\cdot 10^{-5}\,\eta_{\rm
jet/X-ray}$. For this estimate they assume that the jet power is completely
thermalized, and that the energy budget is the same as the X-ray energy release
($\eta_{\rm jet/X-ray}\approx 1$, thus 10 times higher than what we
estimate here). It was also assumed that the thermal energy suffers
from adiabatic cooling due to the Hubble flow, but since radio
galaxies are located in filaments and clusters of galaxies only 1-dimensional 
instead of 3-dimentional expansion was taken into account.

Yamada et al. (1999) use a model for heating and cooling of the
environment of radio lobes. They estimate the jet power from the
black-hole-to-galaxy mass ratio ($f_{\rm bh} = 0.002$) using a
Press-Schechter description of galaxy (and black hole) growth and
assuming a constant fraction ($f_\rg = 0.01)$ of galaxies above
$10^{12}\,M_\odot$ to be active RGs. Since the total energy release in
their description should be higher by an order of magnitude, compared
to what we estimate here, it is not surprising that they get $y = 5.7
\cdot 10^{-5}\, (f_{\rm bh}/0.002)\, (f_{\rm r}/0.01)$, which slightly
violates the present day upper limit.

The present day limit on the Comptonization parameter gives already an
important constraint on the energy budget of RGs, which will be
improved in the near future. If it would be possible to detect the
spectral signature of a rSZ effect, a further major step in the
investigation of radio plasma could be achieved. We do not give
detailed prediction for the rSZ increment, since different to the rSZ
decrement at lower frequencies it strongly depends on the unknown
shape of the electron distribution. But observations of the rSZ effect
in emission could allow a spectral examination of the low-energy end
of the relativistic electron content of the Universe. 

\section{Clusters of Galaxies\label{sec:cluster}}

\subsection{Embedded Radio Plasma\label{sec:RPinCL}}

Clusters of galaxies are known to be strong sources of the thSZ
effect. Since the thermal gas causing the thSZ distortions can also be
observed with X-ray satellites via its bremsstrahlung emission it is
possible to compare the angular diameter with the true line-of-sight
diameter of the cluster. This gives directly the Hubble parameter,
$H_\o$, assuming spherical symmetry at least in a statistical
sense. It is therefore of principal interest to estimate the possible
influence of relativistic plasma on the measured Comptonization.

Radio plasma was mostly produced by outflows from AGN in virialized
cosmological structures such as filaments and clusters of
galaxies. Subsequent flows into deeper gravitational potentials should
have transported a significant fraction of all radio plasma into
clusters of galaxies. If the radio plasma is still unmixed with the
thermal medium, it resembles a cavity in the X-ray emitting ICM
gas. We assume in the following that a (for simplicity constant)
fraction $\Phi_\gh$ of the cluster volume is occupied by radio
ghosts. The X-ray luminosity of this cluster is thus given by
\begin{equation}
\label{eq:LX}
L_X = a_X \,\int \!\! dV\, n_\e^2 \, kT_\e^\frac{1}{2}\, (1-\Phi_\gh) \sim
l_\cl^3\,n_{\e,\o}^2 \, (1-\Phi_\gh)\,.
\end{equation}
$n_\e$ is the cluster electron density, $n_{\e,\o}$ its central value,
and $l_\cl$ the characteristic length scale of the cluster (e.g the core
radius). Since the cluster temperature $kT_\e$ can be determined
spectroscopically we assume it to be known.  The thSZ effect
\begin{equation}
\label{eq:ycluster}
y_\cl = \frac{\sigma_T}{m_\e \, c^2}\, \int \!\!dl  \, n_\e\, kT_\e\,
(1-\Phi_\gh) \sim  l_\cl \, n_{\e,\o}\,
(1-\Phi_\gh) 
\end{equation}
in combination with Eq. (\ref{eq:LX}) allows to measure the typical
scale of the cluster:
\begin{equation}
l_\cl \sim \frac{L_X}{y_\cl^2} \, (1-\Phi_\gh) 
\end{equation}
Since the angular diameter of a cluster should (at least in a
statistical average) be identical to $l_\cl$ the Hubble constant can
be derived:
\begin{equation}
\label{eq:Ho}
H_\o \sim l_\cl^{-1} \sim \frac{y_\cl^2}{L_X \, (1-\Phi_\gh)}\,.
\end{equation}
There are two ways the presence of radio ghosts can affect the
determination of $H_\o$, which shift it in opposite directions: First,
if a significant fraction of the volume is filled with relativistic
plasma, then the true $H_0$ is greater than the one derived under the
assumption that $\Phi_\gh = 0$. Second, if the measured $y_{\rm obs}$
contains a significant contamination due to absorption by the rSZ
effect the true $y_\cl$, and therefore also $H_\o$, is lower.
 
The second effect is negligible for the following reason: The optical
depth of the relativistic electrons of the ghosts can be written as
\begin{equation}
\tau_\gh = \frac{6\,\phi_\gh\,y_\cl}{(1+k_B+k_p)\, \langle \gamma_\e
-1\rangle \,(1-\Phi_\gh)}\,,
\end{equation}
if we assume pressure equilibrium between ghosts and thermal plasma,
and use Eqs. (\ref{eq:epse1}), (\ref{eq:ycluster}), and
\ref{A:eq:Pe3}. Optimistic values for the unknown parameters of the
old radio plasma ($1+k_B+k_p\approx 3$, $\langle \gamma_\e-1\rangle \approx 10$)
give $\tau_\gh \approx 0.2 \,\Phi_\gh\,y_\cl$.  The rSZ decrement is
roughly $-\tau_\gh\, i(x)$, or smaller.  The observed,
rSZ-contaminated $y$-parameter is therefore
\begin{equation}
y_{\rm obs} = y_\cl + \frac{-i(x)}{g(x)} \, \tau_\gh \approx y_\cl
\,(1+0.06\,\phi_\gh )\,.
\end{equation}
In the last approximation we assumed that the measurement is taken
around $x= h\,\nu/kT_\cmb \approx 2$, where the thSZ absorption is at
its maximum and $-i(x)/g(x) \la 0.3$ (see
Fig. \ref{fig:ghij}). Inserting $y_{\rm obs}$ instead of $y_\cl$ into
Eq. (\ref{eq:Ho}) gives an systematic error of $\approx 0.1 \,
\Phi_\gh$, which is an order of magnitude smaller than the error due
to the factor $1-\Phi_\gh$ in the denominator of that equation. The
error due to the rSZ effect could be larger, if the measurement is
done closer to the crossover frequency at $ x= 3.83$. However, this
value of $x$ is inconvenient for detecting the thSZ effect. Also if
$\langle \gamma_\e -1\rangle  \approx 1$ the distortion would be significant, but
this is extremely speculative.

We conclude that the presence of a significant fraction of the volume
of the ICM filled by radio ghosts biases the determination of the
Hubble constant to lower values, if this effect is not accounted for.

\begin{figure}[t]
\begin{center}
\psfig{figure=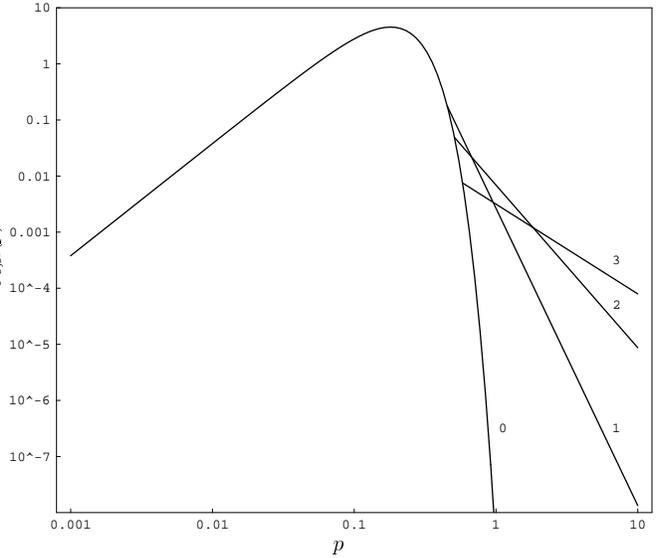,width=0.45 \textwidth,angle=0}
\end{center}
\caption[]{\label{fig:fecoma} The models 0-3 for the electron spectrum
in Coma $f_{\e,{\rm cl}}(p)$.}
\end{figure}

\subsection{Non-Thermal Electrons in the ICM\label{sec:nthe}}
\begin{figure}[t]
\begin{center}
\psfig{figure=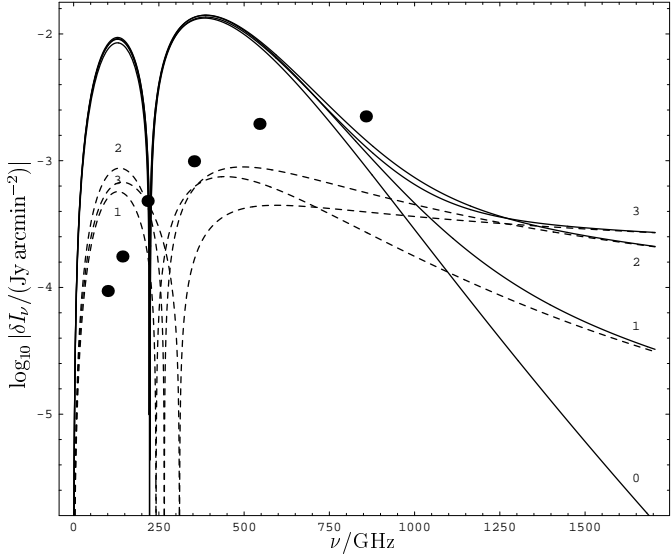,width=0.45 \textwidth,angle=0}
\end{center}
\caption[]{\label{fig:dTTcoma} 
$\log_{10} |\delta I_\nu / ({\rm Jy\,arcmin^{-2}})|$ over $\nu /{\rm
GHz}$ for the center of the Coma cluster. The solid lines are models
0-3 and the dashed lines are the difference of models 1--3 to
model 0. The dots give the frequencies and sensitivities of single
beams of the Planck satellite (Bersanelli et al. 1996; Puget 1998;
see also Tab. \ref{tab:planck}). Since the angular diameter of Coma
(core radius $\hat{=} 10'$) is much larger than the beam of Planck
(FWHM $\le 10'$) a marginal detection (see Appendix B) of a
possible non-thermal electron population producing the HEX excess by
bremsstrahlung can be expected. Balloon experiments or other
high-altitude observatories (at the planned ALMA site or in Antarctica)
will be much more sensitive.  Note that the curves on the lhs
represent absorption, whereas the ones on the rhs represent emission.}
\end{figure}

In some clusters of galaxies a relativistic ICM population of electrons
is visible due to their synchrotron emission, which produces the radio
halos of clusters of galaxies. Also the recently detected extreme
ultraviolet (EUV) excess (Lieu et al. 1996) and the high energy X-ray
(HEX) excess (Fusco-Femiano et al. 1998, 1999) in the Coma cluster
indicate the presence of non-thermal electrons.

Our knowledge about the slope of the electrons spectrum in the ICM is
limited (see En{\ss}lin \& Biermann 1998 for an attempt to compile
the electron spectrum in the Coma cluster). The number density and
therefore the optical depth of the higher energy electrons is too
small to detect the rSZ decrement produced by them. 
See Birkinshaw (1999)\nocite{birkinshaw} for a discussion of the rSZ
decrement from radio emitting electrons and McKinnon et al. (1990) for an
attempt to detect this decrement in radio galaxies.
The rSZ effect emission from these electrons may already have been
observed in the EUV (Hwang 1997, En{\ss}lin \& Biermann 1998, Sarazin
\& Lieu 1998, Bowyer \& Bergh\"ofer 1998) or in the HEX excess
(Fusco-Femiano et al. 1998, 1999). The latter would require a high
number of electrons in the range of $3-5$ GeV, which would
over-produce radio emission if the magnetic fields are as high as
Faraday rotation measurements indicate ($2-10 \,\mu$G; Crusius-Waetzel
et al. 1990; Kim et al. 1991\nocite{kim91}; Feretti et al. 1995, 1999,
Clarke et al. in preparation). Only if fields are as weak as
$0.16\,\mu$G consistency between the expected and observed radio flux
is established (Fusco-Femiano et al. 1998, 1999). As a possible
solution of this apparent contradiction of magnetic field estimates
En{\ss}lin et al. (1999) and Sarazin \& Kempner (2000) proposed that
the HEX excess could alternatively be produced by bremsstrahlung of a
non-thermal high energy tail of the thermal electron
distribution. Such a tail could exist due to in-situ particle
acceleration powered by ICM turbulence (En{\ss}lin et al. 1999; Dogiel
1999; Blasi 2000).

Detection of the IC emission of such a non-thermal tail could confirm
the bremsstrahlung-scenario. In the following we estimate the expected
spectral distortions for the Coma cluster of galaxies.

The electron density of the cluster is assumed to follow a
beta-profile:
\begin{equation}
n_\e(r) = n_{\e,\o} \,\left[ 1+ (r/r_\cl)^2 \right]^{-\frac{3}{2} \,
\beta_\cl}\,,
\end{equation}
where $r_\cl = 400\, {\rm kpc\, h_{50}^{-1}}$ is the core radius,
$\beta_\cl = 0.8$ is the beta-parameter, and $n_{\e,\o} = 2.89\cdot
10^{-3}\,{\rm cm^{-3}}$ is the central thermal electron density (Briel
et al. 1992). We ignore the possible complication due to embedded old
radio plasma discussed in Sect. \ref{sec:RPinCL} for simplicity.

The optical depth of a line of sight passing the cluster center at a
distance $R$ is then
\begin{equation}
\tau_\cl (R) =
{\rm B} \left(\frac{1}{2},\frac{3\,\beta_\cl - 1}{2}\right) \,
 \frac{n_{\e,\o} \,r_\cl \, \sigma_T}{\left[ 1+
(R/r_\cl)^2 \right]^{\frac{3}{2}\,\beta_\cl -\frac{1}{2}}}\,,
\end{equation}
giving a central optical depth of $\tau_\cl (0) = 5.95 \cdot
10^{-3}$. The CMB-distortions are 
\begin{equation}
\delta I_\nu = i_\o\,\delta i(x)= i_\o\, \tau_\cl \,( j(x)-i(x)))\,.
\end{equation}
The electron spectrum is given for a thermal spectrum by
Eq. (\ref{eq:fth}) and for a modified thermal spectrum by Eq.
(\ref{eq:fe,pow}). The non-thermal electrons are treated as an
additional population, increasing slightly the total electron number,
but leaving the number densities of electrons with $p<p_1$
unchanged. We estimate the distortions for a pure thermal spectrum
(model 0) and for three modified thermal spectra, which are able to
explain the HEX excess of Coma by bremsstrahlung (see Fig. 6 in
En{\ss}lin et al. 1999):
\begin{tabular}{llll}
model 1: & $\alpha = 5.3$, & $p_1 = 0.45$, & $E_{\rm
kin,1} = 50 \,{\rm keV}$;\\
model 2: & $\alpha = 2.9$, & $p_1 = 0.51$, & $E_{\rm
kin,1} = 62.5 \,{\rm keV}$;\\
model 3: & $\alpha = 1.6$, & $p_1 = 0.58$, & $E_{\rm 
kin,1} = 80 \,{\rm keV}$;
\end{tabular}
$p_2 = 10$ in all three models. The electron spectra are shown in
Fig. \ref{fig:fecoma} and the resulting CMB distortions at the center
of the Coma cluster in Fig. \ref{fig:dTTcoma}. The differences between
the thermal and the modified thermal models are considerable and
change as a function of frequency, so that future multichannel CMB
telescopes such as high-altitude ground based observatories (e.g. at
the ALMA site or in Antarctica), balloon-experiments, or marginally
even the Planck satellite (see Appendix B) can discriminate
between these models. A detection of this non-thermal SZ effect would
prove in-situ acceleration processes acting in the ICM of Coma. It
would confirm the bremsstrahlung origin of the HEX excess and
therefore solve the discrepancy between Faraday and IC based magnetic
field estimates in Coma.

Since the SZ effect does not depend on distance (as long as the
instrument beam resolves the cluster core) many other clusters can
be investigated for the presence of a non-thermal SZ effects with
future CMB experiments.

\section{Conclusion\label{sec:conclusion}}
\begin{table*}
\begin{tabular}{|l|c|c|c|c|c|c|c|}
\hline
Channel  $k$                & &1&2&3&4&5&6\\
Frequency $\nu_k$ &GHz   & 100 & 143 & 217 & 353 & 545 & 857 \\
Frequency $x_k$            &      &1.76&2.52&3.82&6.21&9.59&15.1\\
Beam FWHM$_k$         &arcmin& 10.7& 8.0 & 5.5 & 5.0 & 5.0 & 5.0\\
Sensitivity $\Delta T/T$&$\mu K/K$&1.7& 2.0 & 4.3 &14.4 &147  &6670\\
Sensitivity $\Delta i(x_k)$& $10^{-6}$&4.09& 7.66& 21.0&43.2&85.2&97.6\\ 
\hline
\end{tabular}
\caption[]{\label{tab:planck}The observing frequencies and
sensitivities of the Planck satellite (Puget
1998)\nocite{puget98}. Note that the sensitivity $\Delta T/T$ refers
to fluctuations in the radiation temperature, rather than to the
system temperature of the receivers.}
\end{table*}

We investigated the transrelativistic Thomson scattering of photons on
a isotropic electron distribution in the optically thin limit
using the photon redistribution kernel (Eq. (\ref{eq:myP})). We
derived for the first time an analytic formula for the scattering by a
population of electrons with a power law momentum distribution,
Eq. (\ref{eq:fe,pow}). We demonstrated that relativistic electron
populations produce a decrement in the cosmic microwave background,
similar to that an absorber with an optical depth of the Thomson depth
of the relativistic electrons would produce. Our formalism can be
generalized to the optical finite case using the techniques described
in Rephaeli (1995a,b) or Molnar \& Birkinshaw (1999).

We applied this theory to radio galaxies and clusters of galaxies:

Although a single radio galaxy produces a negligible SZ decrement, the
combined effect of several radio galaxies and their remnants might
produce a detectable signal. In order to show this we
estimated the total cosmological release in jet power of radio
galaxies using the observed radio luminosity function converted by an
empirical jet power-radio luminosity relation. It is roughly $3\cdot
10^{66} \,{\rm erg\, Gpc^{-3}}\,(f_{\rm power}/3)$, where $f_{\rm
power}$ gives the poorly constrained ratio between true energy content
of radio lobes and the minimum energy estimate. If completely
thermalized this energy would lead to a Comptonization parameter of $y
\approx 6\cdot 10^{-6}$, close to the present day observational limit
of $y < 1.5\cdot10^{-5}$. We argued that roughly $3/4$ of the released
energy remains as a relativistic plasma, which rapidly becomes
unobservable after the activity of the galaxy stopped. Thus the
Comptonization due to heating by radio plasma is expected to be $y
\approx 1.5\cdot 10^{-6}$.

Patches of old plasma, called `radio ghosts', are expected to survive
turbulent erosion unmixed with the ambient medium, so that they are
able to retain a low energy relativistic electron population. The
optical depth for Thomson scattering by radio ghosts is $\tau_\gh
\approx 10^{-7}$ for very optimistic assumptions about the low energy
cutoff of the fresh electrons during injection. Otherwise $\tau_\gh$
is much lower .

If radio ghosts occupy a significant volume in clusters of galaxies,
they affect the determination of the Hubble constant via SZ and X-ray
measurements. The geometric effect of the cavities formed by ghosts in
the intra-cluster gas overwhelms the SZ decrement expected due to
up-scattering of CMB photons by the ghosts' relativistic electron
populations. SZ based $H_\o$ determinations have to be corrected to
higher values due to this effect.

Finally, we demonstrated that future CMB telescopes such as the Planck
satellite seemed to be useful tools to measure supra-thermal electron
populations in clusters of galaxies. These are expected in the case of
turbulent in-situ particle acceleration, and supported by the recently
detected high energy X-ray excess in the Coma and Abell 2556
cluster. Such a detection would be crucial for our understanding of
particle acceleration processes in clusters.

\begin{acknowledgements}
We thank Rashid Sunyaev, Matthias Bartelmann, and Mark Birkinshaw, the
referee, for comments on the manuscript.
\end{acknowledgements}

\appendix
\section{Energy Density and Pressure of a Trans-Relativistic 
Power-Law Electron Distribution\label{app:1}}

If an electron population, with number density $n_{\e,\CR}$, has a
momentum spectrum which is a single power-law (Eq. (\ref{eq:fe,pow}))
the kinetic energy density $\eps_\e$ is
\begin{eqnarray}
\label{A:eq:epse1}
\eps_\e &=&  n_{\e,\CR}\,\langle \gamma_\e -1\rangle  \, m_\e\,c^2 \\ 
&=& n_{\e,\CR} \, \int_0^\infty \!\!\!\! dp\,
f_\e(p)\,(\sqrt{1+p^2}-1)\,m_\e \,c^2 \\
\label{A:eq:Ee}
&=& \frac{n_{\e,\CR} \,m_\e \,c^2}{[p^{1-\alpha}]^{p_1}_{p_2}}\,
\left[\frac{1}{2} \,{\rm B}_\frac{1}{1+p^2} \left(\frac{\alpha
-2}{2},\frac{3 -\alpha}{2} \right) \right.\nonumber\\ 
&& + \left. p^{1-\alpha} \,\left(\sqrt{1+p^2}
-1\right)\,\right]_{p_2}^{p_1}\\
&\approx& \frac{\alpha -1}{\alpha -2}
\frac{[p^{2-\alpha}]^{p_1}_{p_2}}{[p^{1-\alpha}]^{p_1}_{p_2}}\,
{n_{\e,\CR}
\,m_\e\, c^2}\\
&\approx& \frac{\alpha -1}{\alpha -2} \,{n_{\e,\CR} \,m_\e\, c^2\, p_1}
\end{eqnarray}\.
Here, we have used the short-hand notation defined in
Eq. (\ref{eq:short}). The first approximation assumes the particles to
be ultra-relativistic ($1\ll p_1 < p_2$), and the second that the
distribution is dominated by the lower end ($p_1 \ll p_2$ and $\alpha
>2$). For an ultra-relativistic population of electrons (and similar
for protons) the pressure is $P_\e = \frac{1}{3} \, \eps_\e$, but this
is not correct for a transrelativistic population. There
\begin{eqnarray}
\label{A:eq:Pe1}
P_\e &=& n_{\e,\CR} \, \int_0^\infty \!\!\!\! dp\,
f_\e(p)\,\frac{1}{3} \,p\,v(p) \,m_\e \,c \\
\label{A:eq:Pe}
&=& \frac{n_{\e,\CR} \,m_\e \,c^2\,(\alpha
-1)}{6\,[p^{1-\alpha}]^{p_1}_{p_2}}\, \left[{\rm B}_\frac{1}{1+p^2}
\left(\frac{\alpha -2}{2},\frac{3 -\alpha}{2} \right)
\right]_{p_2}^{p_1}\\
\label{A:eq:Pe3}
&\approx& \frac{1}{3} \,\frac{\alpha -1}{\alpha - 2}\,
\frac{[p^{2-\alpha}]^{p_1}_{p_2}}{[p^{1-\alpha}]^{p_1}_{p_2}}\,{n_{\e,\CR}
\,m_\e \,c^2} \approx \frac13 \,\eps_\e
\end{eqnarray}
The ultra-relativistic approximation was again applied, assuming $1\ll
p_1 <p_2$. If this is not justified, correct results  can easily be obtained
from Eq. (\ref{A:eq:Ee}) and (\ref{A:eq:Pe}). 
\section{Detectability of the Cluster-rSZ-Effect with Multifrequency
Experiments\label{app:2}}

In order to demonstrate that the predicted non-thermal CMB distortions
in the clusters like the Coma cluster (Sect. \ref{sec:nthe}) are
detectable with multifrequency experiments we estimate the expected
sensitivity of the Planck satelitte. The spectral distortions at the
location of the observed cluster are produced by temperature
fluctuations of the CMB on the scale of the cluster ($\delta T/T
\approx  10^{-6}...10^{-5}$), the thSZ and kSZ effect of the ICM gas,
and the rSZ effect of the non-thermal electron distribution. The
resulting distortion are therefore
\begin{equation}
\delta i(x) = \sum_{j=1}^3 \tau_j \, f_j (x),
\end{equation}
with 
\begin{eqnarray}
&&\tau_1 = \tau_{\th} ,\,\,\,f_1(x) = j_\th (x) - i(x),\nonumber \\ 
&&\tau_2 =
\tau_{\rel} ,\,\,\,f_2(x) = j_\rel (x) - i(x),\nonumber \\ &&\tau_3 = \delta T/T +
\bar{\beta}_\gas ,\,\,\,f_3(x) = h(x)\,.\nonumber 
\end{eqnarray}
Here, $j_\rel (x)$ are the spectral distorions produced by the
electrons in the non-thermal tail with optical thickness $\tau_\rel$
and spectrum given by $f_{\rm e,
th\&cr}(p;\beta_\th,\alpha,p_1,p_2)-f_{\rm e, th}(p;\beta_\th)$ (with
proper normalization). 

The observing frequencies and sensitivities of the channels ($k$) of
the Planck satelitte are taken from Puget (1998)\nocite{puget98} and
given in Tab. \ref{tab:planck}. We assume the beams to be Gaussian,
and therefore to have the areas $A_{k} = 1.13 \,{\rm FWHM}_k^2$. The
core region of the Coma cluster is $A_{\rm Coma} \approx \pi \, r_{\rm
core}^2 \approx 300\,{\rm arcmin}^2$. Thus, for each frequency 
$N_k = A_{\rm Coma}/A_k$ independent beams probe the CMB. Using a $\chi^2$
analysis allows to dissentangle the different spectral distortions if
sufficient multichannel sensitivity is given. The accuracy in the
determination of $\tau_j$ is given by
$\Delta \tau_j = \sqrt{C_{jj}}$, where $C = A^{-1}$ and
\begin{equation}
A_{jl} = \sum_k \, N_k \, \frac{f_j(x_k)\, f_l(x_k)}{(\Delta i(x_k))^2}
\end{equation}
(Press et al. 1992). Assuming $\tau_\th = 5.95\cdot 10^{-3}$ and
$\tau_\rel$ and $j_\rel$ according to the models 1,2, and 3 (see
Sect. \ref{sec:nthe}) we find that $\tau_\rel / \Delta \tau_\rel =$
0.61, 1.15, 0.96 correspondingly.  This demonstrates that the Planck
mission is expected to give a marginal (1-sigma) detection of such a
supra-thermal electron population in the Coma cluster. Since some
other clusters are also expected to contain supra-thermal electrons --
e.g.  Abell 2256 revealed very recently a similar HEX excess
(Fusco-Femiano 2000)\nocite{2000ApJ...534L...7F} -- a combined signal
from several (HEX excess selected, or merger and post-merger) clusters
might be detectable with statistical significance by Planck. This
demonstrates the ability of future sensitive multichannel CMB
telescopes -- e.g. dedicated baloon experiments -- to detect
supra-thermal electron populations in clusters of galaxies.

\end{document}